\documentclass[aps,prl,showpacs,preprintnumbers,amsmath,amssymb]{revtex4}

\usepackage{epsfig}
\usepackage{epstopdf}
\usepackage{amsmath}
\usepackage{amssymb}
\usepackage{float}
\usepackage[usenames]{color}
\usepackage{color,epsfig,amsmath,amssymb,fancyhdr}

\newcommand{\beq}{\begin{equation}}
\newcommand{\eeq}{\end{equation}}
\newcommand{\ben}{\begin{eqnarray}}
\newcommand{\een}{\end{eqnarray}}
\newcommand{\bi}{\begin{itemize}}
\newcommand{\ei}{\end{itemize}}

\newcommand{\ie}{\textit{i.e.}}

\newcommand{\lambdad}{{\ifmmode \lambda_{\rm d} \else $\lambda_{\rm d}$\fi}}
\newcommand{\pbar}{{\ifmmode \overline{p} \else $\overline{p}$\fi}}
\newcommand{\dbar}{{\ifmmode \overline{d} \else $\overline{d}$\fi}}
\newcommand{\gtilde}{{\ifmmode \tilde{\cal G} \else $\tilde{{\cal G}}$\fi}}
\newcommand{\gtildepos}{{\ifmmode \gtilde^{\; e^+} \else 
    $\gtilde^{\; e^+}$\fi}}
\newcommand{\gtildepbar}{{\ifmmode \gtilde^{\; \pbar} \else 
    $\gtilde^{\; \pbar}$\fi}}

\newcommand{\vmax}{V_{\rm max}}

\newcommand{\kms}{\,{\rm km\,s^{-1}}}

\newcommand{\Msol}{{\ifmmode M_{\odot} \else $M_{\odot}$\fi}}
\newcommand{\Rsol}{{\ifmmode R_{\odot} \else $R_{\odot}$\fi}}
\newcommand{\rhosol}{{\ifmmode \rho_{\odot} \else $\rho_{\odot}$\fi}}
\newcommand{\fsol}{{\ifmmode f_{\odot} \else $f_{\odot}$\fi}}
\newcommand{\gev}{{\ifmmode {\rm GeV} \else ${\rm GeV}$\fi}}

\newcommand{\lcdm}{{\ifmmode \Lambda{\rm CDM} \else $\Lambda{\rm CDM}$\fi}}
\newcommand{\lsim}{\mathrel{\mathop{\kern 0pt \rlap
  {\raise.2ex\hbox{$<$}}}
  \lower.9ex\hbox{\kern-.190em $\sim$}}}
\newcommand{\gsim}{\mathrel{\mathop{\kern 0pt \rlap
  {\raise.2ex\hbox{$>$}}}
  \lower.9ex\hbox{\kern-.190em $\sim$}}}

\begin{document}
\preprint{LAPTH-1274/08 IRFU-09-23}

\title{Cosmic ray lepton puzzle in the light of cosmological N-body simulations}

\author{Pierre Brun}
\affiliation{CEA, Irfu, Service de Physique des Particules, Centre de Saclay, F-91191
Gif-sur-Yvette | France}

\author{Timur Delahaye}
\affiliation{LAPTH, Universit\'e de Savoie, CNRS, B.P.110 74941 Annecy-le-Vieux, France}
\affiliation{ Dipartimento di Fisica Teorica, Universit\`a di Torino / INFN - Sezione di Torino,
Via P. Giuria 1, 10122 Torino | Italy}

\author{J\"urg Diemand}
\affiliation{Hubble Fellow, UCSC, Department of Astronomy and Astrophysics, 1156 High Street, Santa Cruz CA 95064 | USA}

\author{Stefano Profumo}
\affiliation{Santa Cruz Institute for Particle Physics and Department of Physics,
University of California, Santa Cruz, CA 95064 | USA}

\author{Pierre Salati}
\affiliation{LAPTH, Universit\'e de Savoie, CNRS, B.P.110 74941 Annecy-le-Vieux | France}


\begin{abstract}
The PAMELA, ATIC and Fermi collaborations have recently reported an excess in the cosmic
ray positron and electron fluxes. These lepton anomalies might be related to cold dark
matter (CDM) particles annihilating within a nearby dark matter clump. We outline regions of the
parameter space for both the dark matter subhalo and particle model, where data from the different
experiments are reproduced. We then confront this interpretation of the data with the results of the cosmological
N-body simulation Via Lactea II. Having a sizable clump ($\vmax = 9 \, \kms$) at a
distance of only 1.2 kpc could explain the PAMELA excess, but such a configuration
has a probability of only 0.37 percent. Reproducing also the ATIC
bump would require a very large, nearby subhalo, which is extremely unlikely
($p \simeq 3 \times 10^{-5}$). It is even less probable for the smaller Fermi bump to be caused by the presence of such an object. In either case, we predict Fermi will detect the gamma-ray emission from the subhalo. We conclude that under canonical assumptions, the cosmic ray lepton anomalies are
unlikely to originate from a nearby CDM subhalo.
\end{abstract}
%

\maketitle


\vskip -0.25cm

Recent measurements~\cite{PAMELA_frac,ATIC,Fermi} of cosmic ray positrons and electrons
(hereafter dubbed leptons) have stirred a lot of interest, since they may be the first
indirect hint of the presence of particle Dark Matter (DM) in the halo of the Milky Way
(MW)~\cite{Cirelli:2008pk,Bergstrom:2008gr,Cholis:2008hb,Donato:2008jk}.
In the standard cosmic ray picture, positrons are secondary species produced by the
spallation of cosmic ray protons and helium nuclei on the interstellar
medium~\cite{Delahaye:2008ua}. But secondary positrons quite clearly fail to reproduce
the PAMELA~\cite{PAMELA_frac}, ATIC~\cite{ATIC} or  Fermi~\cite{Fermi} measurements. Note that H.E.S.S. also measured cosmic ray leptons spectrum\cite{Collaboration:2008aaa, Aharonian:2009ah}, but with higher threshold, making these measurement less relevant for dark matter interpretations. While
astrophysical primary positron sources exist (the most obvious class of candidates being
pulsars) and can account for the recent measured
anomalies~\cite{Hooper:2008kg,Yuksel:2008rf,Profumo:2008ms},
a more exotic origin is possible. In particular, the observed excess could be sourced
by the annihilations of the massive and weakly interacting particles (WIMPs) proposed to explain
the astronomical DM. Estimates based on standard assumptions on the annihilation
cross-section and on the DM halo fail, however, to reproduce the measured cosmic ray
lepton anomalies. Enforcing a standard thermal DM production in the early universe sets
the pair-annihilation rate a couple of orders of magnitude below what is needed to explain
the data. This issue is usually tackled by assuming {\it boost factors} which can arise
from different arguments.

A first possibility lies in the existence of a particle physics boost factor which
increases the annihilation rate as a result of a mechanism evading the relic density
constraint (e.g. velocity-dependent annihilation cross-sections, non-thermal primordial
production).
Alternatively, an effective astrophysical boost factor could stem from the {\em clumpiness}
of the Galactic DM halo. Indeed, the annihilation rate being proportional to the squared
DM density $\rho^2$, the ``clumpier'' the halo, the higher the signal in a
$\sim \langle \rho^2 \rangle / \langle \rho \rangle^2$ proportion. Although the boost
factor is actually sensitive to the energy~\cite{lavalle:2006vb,brun:2007tn,lavalle:1900wn},
most analyses of the cosmic ray lepton anomalies have so far treated clumpiness either assuming
an energy-independent enhancement of the signal or modifying its spectral shape regardless
of the overall normalization. Moreover, the various contributions to the positron flux
at the Earth have not been derived withxin the same cosmic ray propagation model and are
in general inconsistent with each other.

In this letter, we re-assess~\cite{Cumberbatch:2006tq,Hooper:2008kv,Bringmann:2009ip} the possibility that a single nearby DM clump contributes substantially to the lepton anomalies. Particular attention is paid to cosmic ray propagation. We adjust the clump distance D and luminosity L in order to reproduce the PAMELA, ATIC and Fermi data. The probabilities of these clump configurations are calculated based on the cosmological N-body simulation Via Lactea II~\cite{vialactea}, which allows us to quantify for the first time how unlikely a sufficiently bright nearby CDM subhalo is. We eventually comment about the interplay between the DM particle properties and the clump parameters. As an illustration, we point out extreme configurations where these mutual effects lead to subtle, but relevant modulations of the DM signal.

\label{sec:howto}
%

%
%
The lepton density $\psi = dn/dE$ is related to its source $q$ through
the stationnary cosmic ray diffusion equation described in~\cite{Delahaye:2008ua}.
Lepton propagation throughout the diffusive halo is dominated by space diffusion
and energy losses via synchrotron emission and inverse Compton scattering on the
CMB and stellar light. The diffusion equation is solved in the framework
of the two-zone diffusion model detailed in~\cite{2001ApJ...555..585M} and yields the
flux $\phi = {c \, \psi}/{4 \pi}$. The propagation parameters best-fit the B/C ratio and
correspond to model med of~\cite{2004PhRvD..69f3501D}. A value of $7 \times 10^{15}$ sec is taken for the energy loss timescale.

%
%
The positron flux at the Earth
$\phi_{e^{+}} = \phi_{e^{+}}^{\rm sec} + \phi_{e^{+}}^{\rm s} + \phi_{e^{+}}^{\rm c}$
results here from three contributions.
The astrophysical background $\phi_{e^{+}}^{\rm sec}$ is provided by the secondary
species produced by primary cosmic rays impinging on the interstellar material and
is computed as in~\cite{Delahaye:2008ua}.
The smooth DM halo contributes the source term
\begin{equation}
q_{\rm DM}^{\rm s}({\mathbf x} , E) \; = \;
\frac{1}{2} \, {\left\langle \sigma v \right\rangle} \,
\left\{ {\displaystyle \frac{\rho_{\rm s}({\mathbf x})}{m_{\chi}}} \right\}^{2}
f(E) \;\; ,
\label{source_DM_smooth}
\end{equation}
where $f(E)$ is the energy spectrum of the positrons created in the annihilation
process. The Galactic DM halo density $\rho_{\rm s}$ is borrowed from the
results of the Via Lactea-II simulation, with a spherical profile
featuring an inner (outer) logarithmic slope of -1.24 (-3) and a 28.1 kpc scale parameter.
The local density $\rho_{\odot}$ is equal to 0.37 GeV cm$^{-3}$. The galactocentric
distance of the Earth is $r_{\odot} = 8.5$ kpc. 
%
Finally, the contribution of a nearby clump located at ${\mathbf x_c}$ can be
expressed as
\begin{equation}
q_{\rm DM}^{\rm c}({\mathbf x} , E) \; = \;
\frac{1}{2} \; {\left\langle \sigma v \right\rangle} \;
 {\displaystyle \frac{L}{m_{\chi}^{2}}} \;
\delta^{3}({\mathbf x} - {\mathbf x_{c}}) \; f(E) \;\; ,
\label{source_DM_clump}
\end{equation}
where $L = \int_{\rm clump} \, \rho_{c}^{2}({\mathbf x}) \, \text{d}^3{\mathbf x}$
is defined as the subhalo luminosity. 
Furthermore, as clumps are treated here as point-like objects
\footnote{The scale radii of the clumps are always much smaller than the typical lepton
diffusion length.},
the source terms $q_{\rm DM}^{\rm s}$ and $q_{\rm DM}^{\rm c}$ add up directly to yield
the total DM lepton signal.

%
%
The annihilation cross-section of the DM particles under scrutiny is set equal to
$\langle \sigma v \rangle = 3 \times 10^{-26}$ cm$^{3}$ s$^{-1}$. This canonical value
matches a thermal production of DM in the early universe.
We also consider 100 GeV and 1 TeV WIMPs as benchmark cases.
Finally, inspired by~\cite{Cirelli:2008pk,Donato:2008jk}, we concentrate here on
leptophilic species and considered either a pure e$^{+}$e$^{-}$ annihilation final state
(positronic line) or an equal production of charged leptons
e$^{\pm}$ + $\mu^{\pm}$ + $\tau^{\pm}$. Notice that a case for both scenarios can be made
from the model-building perspective.
We also tried to use models featuring pure ${\rm b \bar{b}}$ or W$^{+}$W$^{-}$ annihilation
final state, but we disregard them since the required clump configurations are extremely unlikely ($p< 10^{-6}$).
%

%
\begin{figure*}[t]
\vskip -0.75cm
\centering
\includegraphics[width=7.5cm]{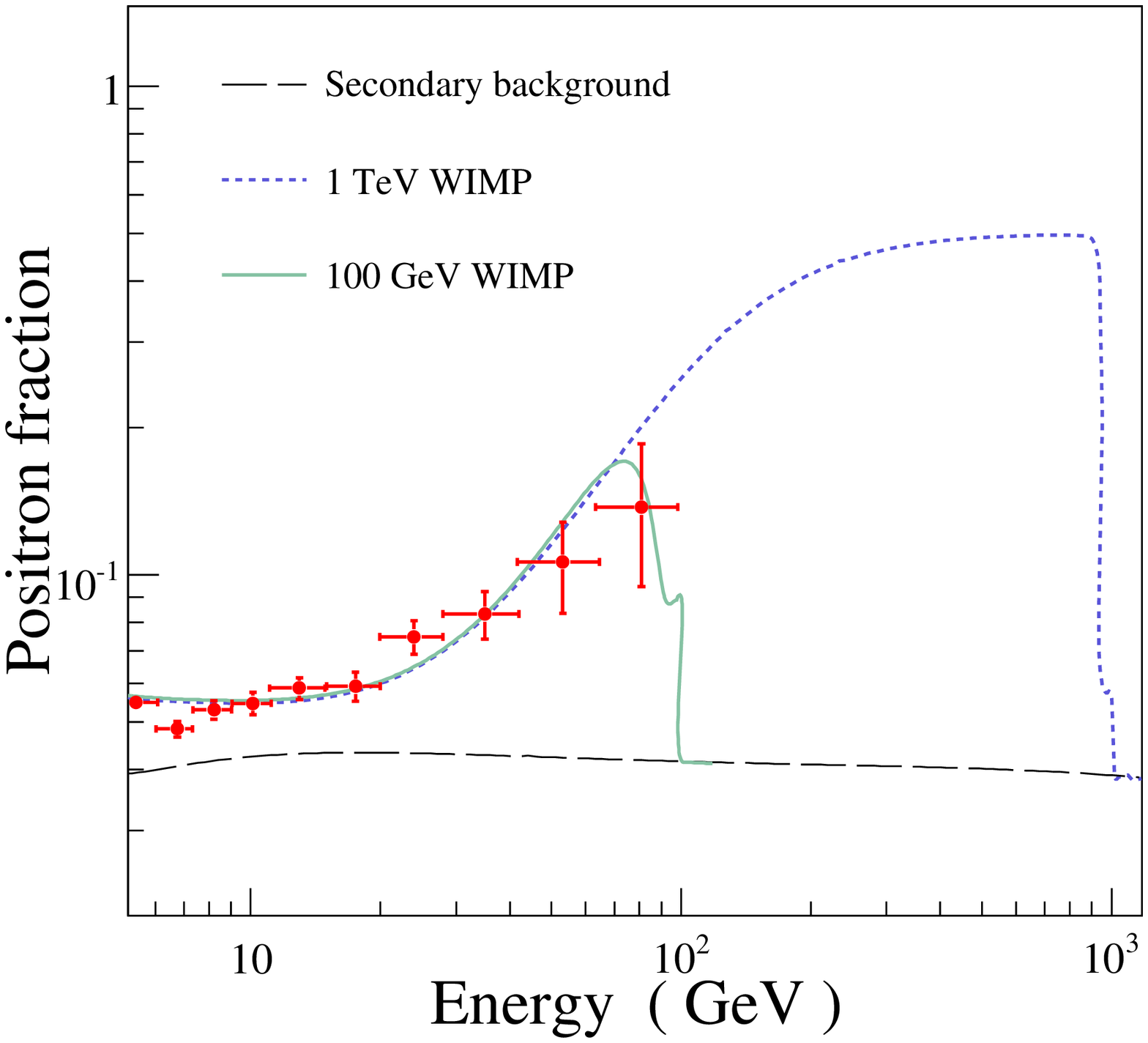}
\includegraphics[width=8.3cm]{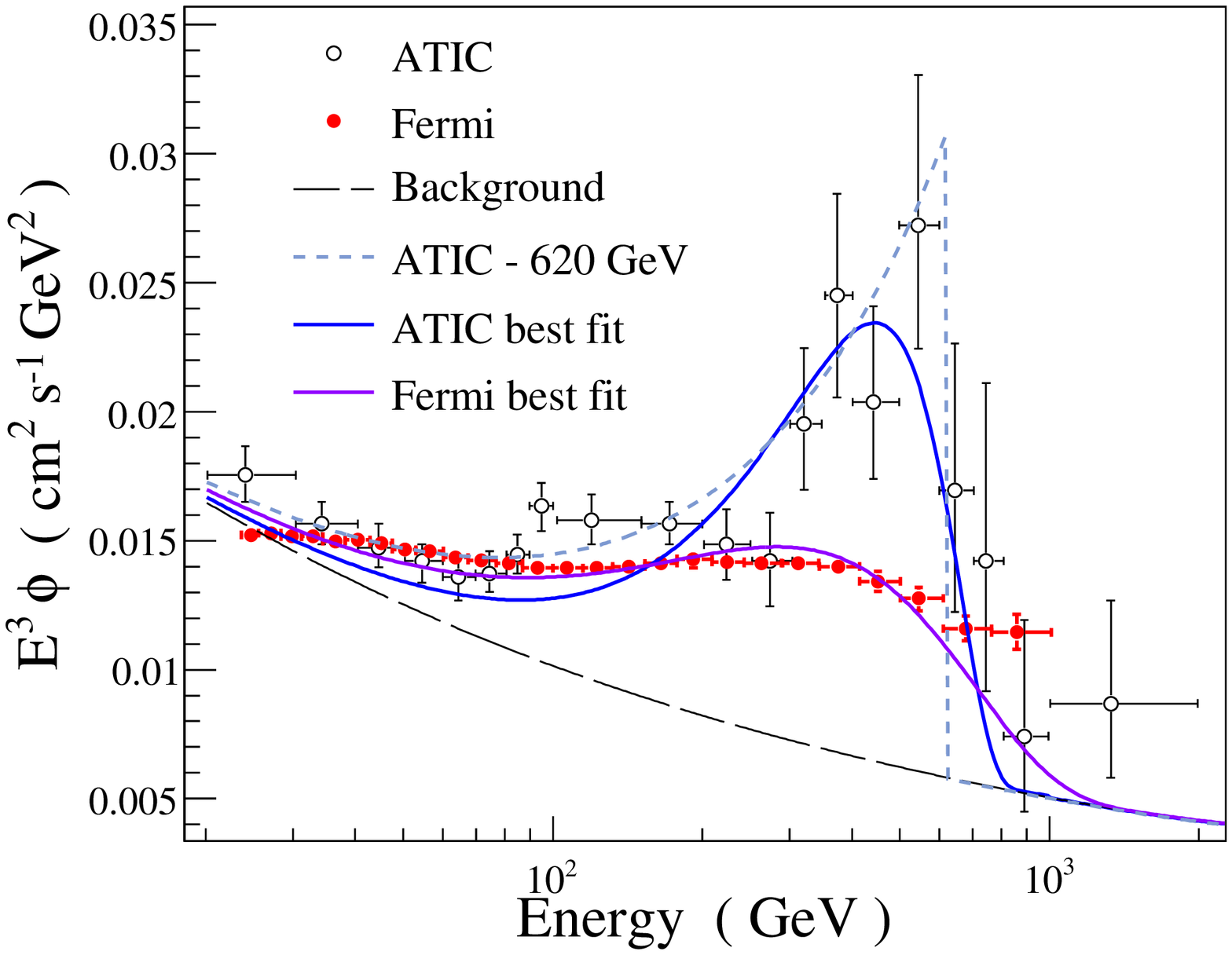}
\caption{
Best fits to the PAMELA data in the case of a positronic line (see the e$^{+}$/e$^{-}$ row of Tab.~\ref{tab:fits}) (left panel) 
and fits to the ATIC data (right panel).
}
\label{fig:fig1}
\vskip -0.5cm
\end{figure*}
%

%
The particle physics framework being set, we perform fits to the PAMELA, ATIC and Fermi
data which include a smooth DM component plus a contribution from a DM subhalo
whose luminosity $L$ and distance $D$ are free parameters.
For PAMELA, we compute the positron fraction
${\phi_{e^{+}}}/{\phi_{e^{+}} + \phi_{e^{-}}}$ where we use for $\phi_{e^{-}}$
the observed cosmic ray electron flux measured by AMS~\cite{Alcaraz:2000bf}
and HEAT~\cite{Beatty:2004cy} and parameterized by Casadei \&
Bindi~\cite{2004ApJ...612..262C}.
As regards ATIC and Fermi, we derive the total lepton flux $\phi_{e^{+}} + \phi_{e^{-}}$,
assuming that the electron background $\phi_{e^{-}}^{\rm back}$ is given at high
energy  by the Casadei \& Bindi fit and adding a DM contribution equal
to $\phi_{e^{+}}^{\rm s} + \phi_{e^{+}}^{\rm c}$.
Solar modulation is implemented using the force field
approximation~\cite{1987A&A...184..119P} with a Fisk potential of 300 MV.
Fig.~\ref{fig:fig1} illustrates the fit results in the case of e$^{\pm}$ direct
production. For PAMELA, both 100 GeV and 1 TeV WIMPs can accommodate the excess.
As far as ATIC is concerned, we reproduce the observed feature in the case of a
1 TeV WIMP if we assume a DM clump with luminosity
$2.98 \times 10^{9}$ M$_{\odot}^{2}$ pc$^{-3}$ lying at a distance of 1.52~kpc
from the Earth. The ATIC excess was reported in~\cite{ATIC} where
it was interpreted as evidence for a 620~GeV Kaluza-Klein species. We confirm
that result in the case of a positronic line. In that case, no satisfying adjustment can
be found adding a nearby subhalo, WIMP annihilations take place only inside a
smooth Galactic DM distribution and the required cross-section
$<\sim 10^{-24}$ cm$^{3}$ s$^{-1}$, {\ie} two orders of magnitude above
our canonical value.As we shall see in the following, the fit to the Fermi data points towards an incredibly bright clump. Indeed, the feature seen by Fermi is not very peaked and quite spread and requires, because of propagation effect, both a very large mass for the DM particle and a quite far away clump.
We therefore exclude the combination of a thermal relic density plus a bright clump as a solution to this puzzle and we do not mention this case when dealing with other messengers. All parameters found in the best-fit cases are displayed in Tab.~\ref{tab:fits}.
%
%
\begin{table}[h]
\centering
\begin{tabular}{|c||c|c||c||c|}
\hline
& \multicolumn{2}{|c||}{PAMELA} & ATIC & Fermi \\
\hline
m$_{\chi}$ (GeV) & 100 & 1 000  & 1 000 & 2500\\
\hline
\hline
e$^{+}$/e$^{-}$ &
$1.22 - 1.07 \!\cdot\! 10^7$ & $0.78 - 3.56 \!\cdot\! 10^9$ & $1.52 - 2.98 \!\cdot\! 10^9$ & $2.68 - 5.53 \!\cdot\! 10^{10}$\\
\hline
e$^{\pm}$ + $\mu^{\pm}$ + $\tau^{\pm}$ &
$0.44 - 2.51 \!\cdot\! 10^7$ & $0.27 - 9.84 \!\cdot\! 10^9$ & $0.25 - 8.78 \!\cdot\! 10^9$ & $2.81 - 2.17 \!\cdot\! 10^{11}$\\
\hline
\end{tabular}
\caption{
Best fit values of the ($D ; L$) couple in units of
$(\text{kpc} ; \text{M}_{\odot}^{2} \, \text{pc}^{-3})$
for various DM particle masses and annihilation channels.}
\label{tab:fits}
\vskip -0.5cm
\end{table}
%

%
\begin{table}[h!]
\footnotesize
\centering
\begin{tabular}{|c||c|c||c|}
\hline
& \multicolumn{2}{|c||}{PAMELA} & ATIC \\
\hline
m$_{\chi}$ (GeV) & 100 & 1 000  & 1 000 \\
\hline
\hline
e$^{+}$/e$^{-}$ &
  $0.23$ (${\rm b \bar{b}}$)     & $0.066$ (W$^{+}$W$^{-}$)       & $0.13$ (W$^{+}$W$^{-}$) \\
\hline
e$^{\pm}$ + $\mu^{\pm}$ + $\tau^{\pm}$ &
  $0.063$ (${\rm b \bar{b}}$)    & $0.0074$ (W$^{+}$W$^{-}$)      & $0.055$ (W$^{+}$W$^{-}$) \\
\hline
\hline
e$^{+}$/e$^{-}$ &
  $0.95$                         & $12.7$                         & $3.9$ \\
\hline
e$^{\pm}$ + $\mu^{\pm}$ + $\tau^{\pm}$ &
  $28.6$                         & $370$                          & $12.9$ \\
\hline
\end{tabular}
\caption{
Upper rows~:
maximal values of the branching ratio $F_{\bar{p}}$ of WIMP annihilation into
${\rm b \bar{b}}$ (100 GeV) or W$^{+}$W$^{-}$ (1~TeV) pairs allowed by the PAMELA
antiproton data~\cite{Adriani:2008zq}.
Lower rows~:
the clump $\gamma$-ray flux above 0.1~GeV is expressed in units of
$3 \times 10^{-9}$ $\gamma$~cm$^{-2}$~s$^{-1}$~\cite{Atwood:2009ez}.}
\label{tab:pbar_gamma}
\vskip -0.3cm
\end{table}
%

%
%
DM annihilations within these clumps not only produce charged leptons but also
antiprotons and $\gamma$-rays.
We first allow a small branching ratio $F_{\bar{p}}$ into antiprotons either through
the ${\rm b \bar{b}}$ channel when the WIMP is light or through the W$^{+}$W$^{-}$
channel for a 1~TeV species. We compute the total antiproton flux
$\phi_{\bar{p}} = \phi_{\bar{p}}^{\rm sec} + \phi_{\bar{p}}^{\rm s} + \phi_{\bar{p}}^{\rm c}$
with the same cosmic ray and DM models as for positrons. Requiring that the resulting signal
does not exceed the PAMELA antiproton data~\cite{Adriani:2008zq} by more than $1\sigma$,
we derive the upper limits on $F_{\bar{p}}$ featured in Tab.~\ref{tab:pbar_gamma}. Antiprotons
are not forbidden if they are produced together with leptons but their abundance in the annihilation debris is sufficiently suppressed.
We also present in Tab.~\ref{tab:pbar_gamma} conservative estimates for the detectability of
the $\gamma$-ray emission from the lepton-fitting clumps of Tab.~\ref{tab:fits}.
The $\gamma$-ray flux at the Earth is expressed in units of Fermi $5\sigma$ sensitivity
over 1~year of data taking for high-latitude point-like sources~\cite{Atwood:2009ez}.
This is probably roughly close to the best the LAT instrument can do for a low-latitude
source over its lifetime. We conclude that in all cases the LAT will detect the clumps.
Notice that one of the leptophilic 1~TeV PAMELA cases has a very large flux of $1.11 \times 10^{-6}$ $\gamma$~cm$^{-2}$~s$^{-1}$ and is already of the order of the EGRET point-source sensitivity of $2 \times 10^{-7}$ $\gamma$~cm$^{-2}$~s$^{-1}$~\cite{1999yCat..21230079H}, and should have been already detected and correspond to one of EGRET unidentified sources.

\begin{figure}[t]
\vskip -0.75cm
\centering
\includegraphics[width=.5\columnwidth]{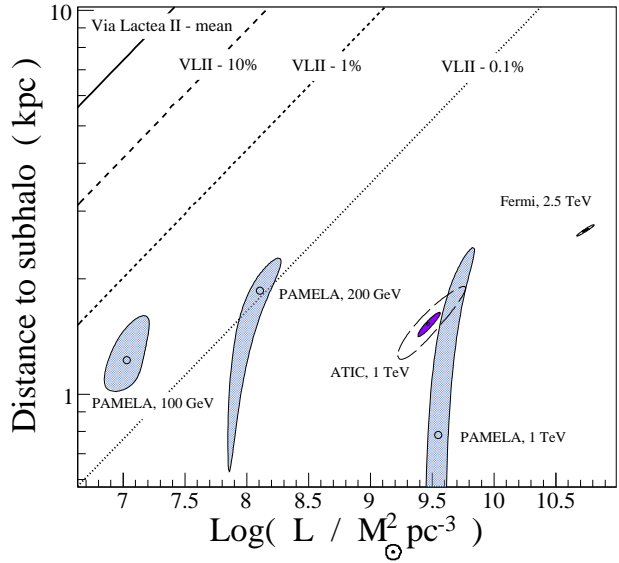}
\caption{
Best fit results in a clump luminosity-distance plane for different configurations,
together with probabilities inferred from Via Lactea-II results.
}
\label{fig:fig2}
\vskip -0.5cm
\end{figure}

Fig.~\ref{fig:fig2} shows the probability of having the nearest DM clump of luminosity $L$
within a distance $D$ from the Sun. The abundance of nearby clumps and their properties are
taken directly from the Via Lactea-II (VL-II) simulation~\cite{vialactea}. The high mass and
force resolution, combined with a physical time step criterion~\cite{Zemp2007}, allow VL-II
to resolve subhalos even in the dense environment near the solar circle. The mean separation of
subhalos with peak circular velocities $\vmax > 5 \, \kms$ is 9.6 kpc and their luminosities are
\begin{equation}
L = 7.91\times 10^5 \; {\rm M}_{\odot}^{2}\, {\rm pc^{-3}} \,
\left( {\displaystyle \frac{\vmax}{5 \, \kms}} \right)^{3}
\sqrt{ {\displaystyle \frac{c_V}{2 \times 10^6}} } .
\end{equation}
For comparison, the smooth VL-II main halo 
has a luminosity of $L = 3.4 \times 10^{9} \; {\rm M}_{\odot}^{2}\, {\rm pc^{-3}}$, while the total luminosity is about 10 times higher~\cite{vialactea}.
At a given $\vmax$ we assume a log-normal distribution of luminosities with factor of
3 scatter, motivated by the substantial variance in the concentration $c_V$ found in nearby
subhalos~\cite{vialactea}. The bold line gives the median distances calculated from
a random sample of observer positions. The long-dashed, dashed and doted lines stand for the 10th
1st and 0.1st percentiles, respectively. The points represent the locations of the best fits
to the data in the $L-D$ plane while the surrounding contours display the $1\sigma$ excursions
around these best fit values (as well as $3\sigma$ for ATIC and Fermi). We find that clumps fitting the PAMELA, ATIC and Fermi data are far
from the natural values indicated by VL-II. The most probable configuration
is the PAMELA fit with a 100 GeV WIMP, which is inside the Via Lactea $3\sigma$ contours.
That configuration is found in 0.37\% of all realizations. However, this scenario
cannot accomodate ATIC data because $m_{\chi}$ is too small. Increasing the mass of the DM
particle requires even brighter and less likely clumps at $D\simeq 1$ kpc.
As illustrated by the different PAMELA fit
contours of Fig.~\ref{fig:fig2}, the parameter degeneracy also increases as $m_{\chi}$ gets
higher.
Basically, the PAMELA measurements do not constrain the spectral shape of the signal
above 100 GeV and leave more lever-arm to the fits when WIMPs are heavy.
For TeV WIMPs, there are clump properties which reproduce both the ATIC and PAMELA excess (see Fig.\ref{fig:fig2}), and such a source would be well within the reach of Fermi (Tab.~\ref{tab:pbar_gamma}). However, in the standard CDM halo it is very unlikely to exist ($p \simeq 3 \times 10^{-5}$). A DM spike or a higher cross-section would be required to get the needed luminosity from a smaller, more probable nearby subhalo.%
%
%
As shown in Tab.~\ref{tab:pbar_gamma}, the corresponding subhalo is within reach of Fermi.
Concerning the best fit to the Fermi data, it points at the need for a clump that should be brighter than the whole Milky Way. We can safely associate a zero probability to this configuration. 
Note finally that the Via Lactea II contours extrapolate at lower values of the distance and clump luminosity. The corresponding clumps are not of particular interest for this particular study as  for mean Via Lactea clumps, the natural luminosity decreases faster than what we gain from placing the clump closer.
%
%


Our investigation of the lepton anomalies in the presence of a nearby DM clump has
finally led us to discover a subbtle interplay between the injection spectrum, the
position of the substructure and the signal at the Earth. In particular, the commonly
used criterion of a sharp cut-off as a smoking-gun signature of DM models producing
leptons ({\it e.g.} in Universal Extra Dimension Kaluza-Klein models~\cite{Hooper:2007qk})
is misleading if most of the DM signal stems from a highly
clumpy halo.  At TeV energies, leptons detected at the Earth
are produced within a distance of $\sim 1$ kpc and a clump becomes less visible if it
lies outside that region.
The opposite case is also possible since one can obtain very peaked lepton
spectra even with a soft injection spectrum, as long as the nearest clump is close
enough. In general, as far as spectral shapes are concerned, it is important to bear
in mind the  spectral distortions induced by cosmic ray propagation.
%
\begin{figure}[h]
\vskip -0.4cm
\centering
\includegraphics[width=.4\columnwidth]{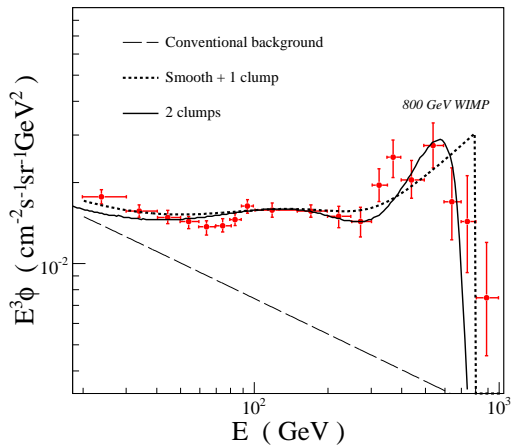}
\vskip -0.2cm
\caption{
Electron-positron spectra resulting from one (dotted) or two (solid) nearby
subhalos. In the latter case, a 620 GeV DM species with thermal annihilation
cross-section is considered.
}
\label{fig:fig3}
\vskip -0.2cm
\end{figure}

As an illustration, we show in Fig.~\ref{fig:fig3} how the signal from DM can
exhibit a double-bump feature if, in addition to the contribution from a smooth
DM halo, two nearby clumps are taken into account (solid light-colored line). In this example,
the subhalos lie at a distance of 0.9 and 4.3 kpc from the Earth and their luminosities are of order
$10^{8}$ and $10^{10}$
${\rm M}_{\odot}^{2}\, {\rm pc^{-3}}$ respectively. Indeed, a specificity of lepton propagation is that regardless of the overall normalization, a feature observed at energy $E$ can always be produced by
a source at distance $D$ with an injection energy $E_{S}$ as long as $D^{2} \propto E^{-0.3} - E_{S}^{-0.3}$
\footnote{The 0.3 exponent depends on the cosmic ray propagation model.}.
However, without additional enhancements, having such bright nearby clumps is practically ruled out and we use them here for pedagogical purpose. CDM subhalos with $L > 10 ^8\, {\rm M_{\odot}^2 pc^{-3}}$ have $\vmax > 25 \, \kms$ and host relatively bright dwarf galaxies~\cite{2007ApJ...671.1135K}. If such dwarfs existed nearby, they should have been observed. 
As regards the light-colored dased curve of the right panel of Fig.~\ref{fig:fig3}, the sharp edge at
800~GeV is associated to a strong local DM annihilation
and is produced by the VL-II smooth halo whereas the bump at $\sim 100$ GeV
comes from a single nearby clump located at
3.2 kpc. The cross section has been increased up to a value of $10^{-23}$ cm$^{3}$ s$^{-1}$.
This case seems somewhat more probable than the 2-clumps configuration.
These examples illustrate how tricky boost factors are.
Shifting upwards the DM cosmic ray fluxes turns out to be wrong especially
in the light of the subbtle effects introduced by propagation.

This letter is focused on the astrophysical boost associated to a nearby DM
clump. We have restrained ourselves from using annihilation cross-sections
in excess of the canonical value of $3 \times 10^{-26}$ cm$^{3}$ s$^{-1}$.
As a final remark, we point out that all the fitted spectra would
have been obtained with significantly less luminous subhalos should the
annihilation cross-section be directly enhanced. If the Sommerfeld
effect~\cite{Hisano:2003ec,MarchRussell:2008tu} is at play, the signal
from small clumps is enhanced with respect to the contribution from larger
substructures~\cite{Lattanzi:2008qa}
since the velocity dispersion of DM particles decreases with
the mass of the host subhalo. The blue contours of Fig.~\ref{fig:fig2} are
shifted towards smaller values of $L$ and get nearer to the mean predictions
of the VL-II simulation, with a much larger probability of occurrence (by simply assuming $L \rightarrow L \times c/\vmax$, the probability of ATIC best fit case increases from $p \simeq 3 \times 10^{-5}$ to 14 percent!).
Due to the lack of recent data, we had to do some assumptions on the
electron spectrum and propagation parameters. Therefore new data may
rule out some of these assumptions and hence lead to different
conclusions.
Future detailed measurements of the lepton spectrum from Fermi, and (on a
much longer time-scale) of the positron fraction from {\it e.g.} PEBS
\cite{Gast:2009yh} will provide us with much greater information on the
nature and origin of the lepton anomalies, and perhaps shed light on
galactic dark matter.

\vskip 0.5cm
\begin{acknowledgments}
T.D. is grateful for financial support from the Rh\^one-Alpes region (Explora'Doc) and from the Internationnal Doctoral school in AstroParticle Physics (IDAPP). J.D. is supported by NASA. is partly supported by US DoE Contract DE-FG02-04ER41268, NASA Grant Number NNX08AV72G and NSF Grant PHY-0757911. This work was partially supported by the French ANR project, {\tt ToolsDMColl} No BLAN07-2-194882.
\end{acknowledgments}

\bibliography{bddps}


\end{document}